\documentclass[conference]{IEEEtran}
\usepackage{cite}

\ifCLASSINFOpdf
  \usepackage[pdftex]{graphicx}
\else
\fi
\usepackage[cmex10]{amsmath}
\usepackage{wasysym}

\begin{document}
%
\title{Generating Entanglement between Atomic Spins with Low-Noise Probing of an Optical Cavity}

\author{\IEEEauthorblockN{Kevin C. Cox, Joshua M. Weiner, Graham P. Greve, and James K. Thompson
\IEEEauthorblockA{JILA and Department of Physics \\
University of Colorado and National Institute of Standards and Technology \\
Boulder, Colorado USA 80309 \\
Email: jkt@jila.colorado.edu}}}


%


\maketitle

\newcommand*{\ket}[1]{\ensuremath{\left|#1\right\rangle}}
\newcommand*{\bra}[1]{\ensuremath{\left\langle#1\right|}}
\newcommand*{\ketup}{\ensuremath{\ket{\uparrow}}}
\newcommand*{\ketdown}{\ensuremath{\ket{\downarrow}}}
\newcommand*{\Rb}{\ensuremath{^{87}\mathrm{Rb}}}
\newcommand*{\FSR}{\ensuremath{f_\mathrm{FSR}}}

\newcommand*{\kv}[1]{\ensuremath{\vec{k}_{#1}}}
\newcommand*{\Nup}{\ensuremath{N_{\uparrow}}}
\newcommand*{\Ndown}{\ensuremath{N_{\downarrow}}}
\newcommand*{\Nupbar}{\ensuremath{\overline{N}_{\uparrow}}}
\newcommand*{\Ndownbar}{\ensuremath{\overline{N}_{\downarrow}}}
\newcommand*{\avg}[1]{\ensuremath{\left\langle#1\right\rangle}}
\newcommand*{\std}[1]{\ensuremath{\Delta\left(#1\right)}}
\newcommand*{\rabi}{\ensuremath{\Omega_\uparrow}}
\newcommand*{\op}{\ensuremath{\omega_+}}
\newcommand*{\om}{\ensuremath{\omega_-}}
\newcommand*{\opm}{\ensuremath{\omega_\pm}}
\newcommand*{\ohf}{\ensuremath{\omega_\mathrm{hf}}}
\newcommand*{\oehf}{\ensuremath{\omega_\mathrm{ehf}}}

\newcommand*{\detmethod}{\ensuremath{\eta_d}}
\newcommand*{\dc}{\ensuremath{\delta_c}}
\newcommand*{\dpr}{\ensuremath{\delta_p}}
\newcommand*{\ms}{\ensuremath{m_s}}
\newcommand*{\msp}{\ensuremath{m_s^\mathrm{proj}}}
\newcommand*{\msopt}{\ensuremath{m_s^\mathrm{opt}}}
\newcommand*{\msbar}{\ensuremath{\overline{m}_s}}

\newcommand*{\dwproj}{\ensuremath{\Delta\omega^\mathrm{proj}}}

\newcommand*{\prepnoise}{\ensuremath{\zeta_\mathrm{prep}}}
\newcommand*{\qpn}{\ensuremath{\Delta J_{z,\mathrm{CSS}}}}

\newcommand*{\J}{\ensuremath{\hat{\bf J}}}
\newcommand*{\R}{\ensuremath{{\cal R}}}
\newcommand*{\sq}{\ensuremath{\xi_m}}
\newcommand*{\sqopt}{\ensuremath{\sq^\mathrm{opt}}}

\begin{abstract}
Atomic projection noise limits the ultimate precision of all atomic sensors, including clocks, inertial sensors, magnetometers, etc.   The independent quantum collapse of $N$ atoms into a definite state (for example spin up or down) leads to an uncertainty $\Delta \theta_{SQL}=1/\sqrt{N}$ in the estimate of the quantum phase accumulated during a Ramsey sequence or its many generalizations.  This phase uncertainty is referred to as the standard quantum limit.  Creating quantum entanglement between the $N$ atoms can allow the atoms to partially cancel each other's quantum noise, leading to reduced noise in the phase estimate below the standard quantum limit.  Recent experiments have demonstrated up to $10$~dB of phase noise reduction relative to the SQL by making collective spin measurements.  This is achieved by trapping laser-cooled Rb atoms in an optical cavity and precisely measuring the shift of the cavity resonance frequency by an amount that depends on the number of atoms in spin up.  Detecting the probe light with high total efficiency reduces excess classical and quantum back-action of the probe.  Here we discuss recent progress and a technique for reducing the relative frequency noise between the probe light and the optical cavity, a key requirement for further advances.  
\end{abstract}


%
\IEEEpeerreviewmaketitle

\section{Introduction}
Atoms and molecules make excellent sensors of time \cite{Bloom14,Hinkley2013}, accelerations \cite{GBK97}, and fields \cite{Kitching2011}, and enable precise tests of fundamental physics \cite{PhysRevLett.98.070801,PhysRevLett.106.080801,ACMEEDM2014}. This is because quantum mechanics provides certainty that the atoms can be made nearly identical using modern optical pumping and laser-cooling and trapping techniques.   The value of the physical quantity to be sensed is most often encoded in the impact of the physical quantity on the rate at which a quantum phase develops  between  quantum states.  Here, we will consider spin states, but quantum phases can also be measured between other types of quantum states, such as momentum states in matter wave interferometers.  The high accuracy provided by quantum mechanics  must be balanced against the fundamental quantum uncertainty that quantum mechanics imposes on the measurement of the evolved phase.  The uncertainty can be mitigated by using many independent atoms $N$ in parallel, as is done in optical lattice clocks or matter wave interferometers.   The independent quantum collapse or projection noise \cite{Itano1993} of each atom is averaged down to an uncertainty in the  estimate of the evolved phase scaling as $\Delta \theta_{SQL} = 1/\sqrt{N}$~radians, a limit known as the standard quantum limit.

\begin{figure}[!t]
\includegraphics[width=3.5in]{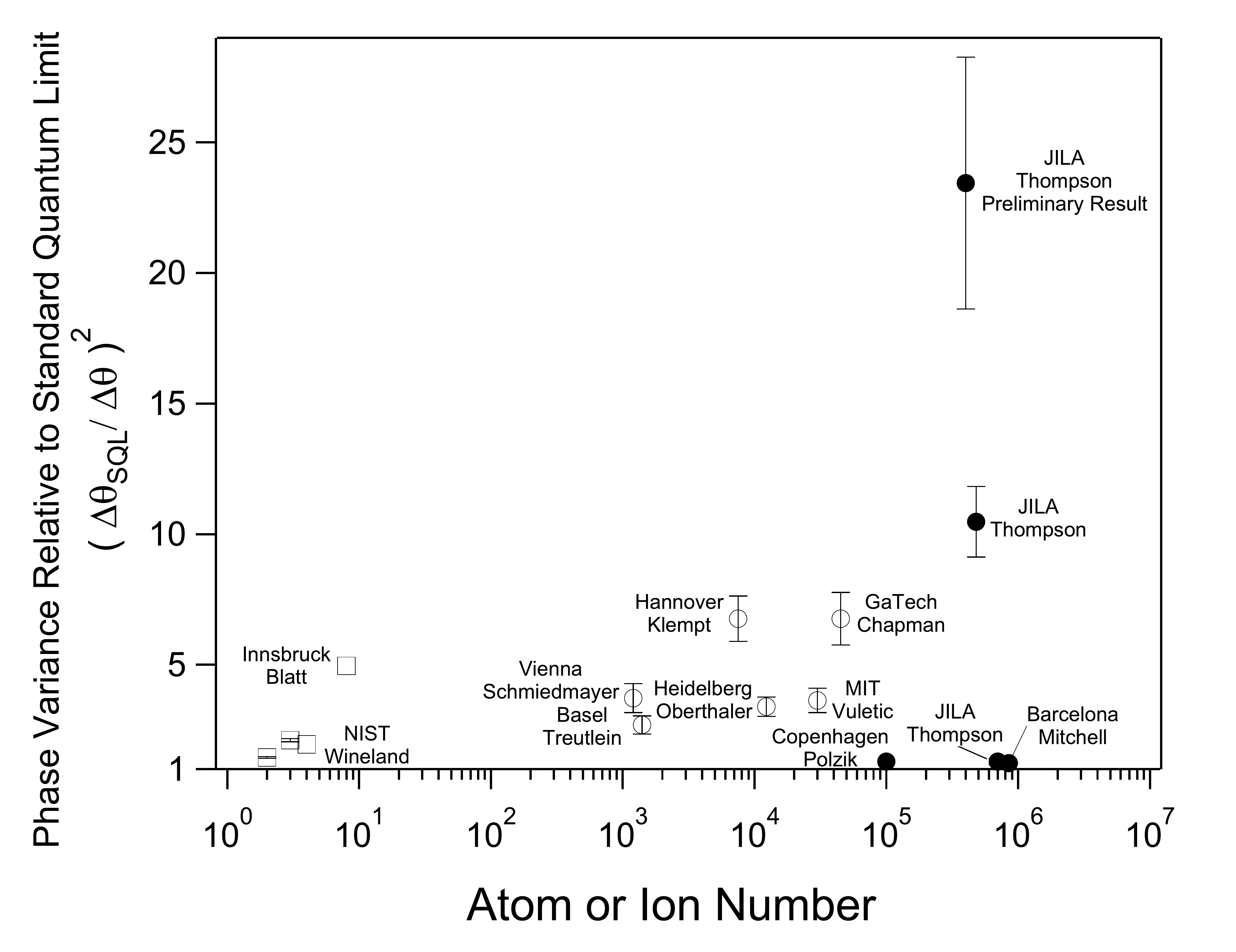}
\centering
\caption{A partial summary of the directly observed entanglement enhanced estimation of a quantum phase relative to the standard quantum limit versus the number of atoms or ions used.  Here, 1 means no enhancement.  Larger inferred values (\textit{i.e.}, with various background subtractions) are reported in many of the references.  The solid circles  $\newmoon$ were achieved using coherence preserving measurements such as described here in either optical cavities \cite{Squeezing_Bohnet_2014,PhysRevLett.106.133601} or in free space \cite{AWO09,SKN12}.  The open circles $\fullmoon$ were achieved using atomic collisions to generate non-linear twisting or parametric type interactions \cite{Bucker2011,Chapman12,Treutlein2013,Klempt2014,Oberthaler2014} or an optical cavity to create effective one-axis twisting interactions \cite{LSM10}.  The open boxes $\Box$ were achieved in ion traps using quantum logic type interactions to generate cat-states \cite{Leibfried2004,Leibfried2005,Monz2011} or squeezed states \cite{Wineland01}.}
\label{figEntanglementSoFar}
\end{figure}

Quantum entanglement between the $N$ atoms can allow the randomness in the measurement-induced collapse of one atom to be partially cancelled by biasing the collapse of other atoms\cite{Wineland1992,Kitagawa1993}.  Theoretically, it is possible to produce entangled states such that quantum projection noise is cancelled by this compensation.  However, the last atom to be measured will not have another atom present to cancel its projection noise.  The ``noise of the last atom" leads to a more fundamental phase estimation uncertainty $\Delta \theta_{HL} \approx 1/N$, known as the Heisenberg limit.  For large numbers of atoms, this enhancement is potentially dramatic.

Small numbers of ions of order 10 or fewer \cite{WinelandNature2000,Wineland01,Leibfried2004,Leibfried2005,Monz2011,Noguchi2012} have achieved phase imprecision close to the Heisenberg limit.  However, it is only recently that larger ensembles of atoms have been entangled and shown to improve phase estimation beyond the standard quantum limit.  The approaches have included twisting operations  using atomic collisions \cite{EGW08,GZN10,Chapman12,Treutlein2013,Oberthaler2014} or an optical cavity \cite{LSM10}, parametric two-mode squeezing driven by atomic collisions \cite{Klempt11,Bucker2011,Klempt2014},  and coherence preserving collective measurements of atoms in optical cavities \cite{SLV10,Leroux10,PhysRevLett.106.133601,Squeezing_Bohnet_2014} and in free space \cite{AWO09,Wasilewski2010,SKN12}.  Figure~\ref{figEntanglementSoFar} attempts to partially summarize the observed enhancements in phase estimation sensitivity relative to the standard quantum limit.  The results are plotted versus atom or ion number since the rms phase estimation sensitivity of the standard quantum limit scales as $1/\sqrt{N}$.

One heuristic way to think about the importance of scaling entanglement to larger atom number is to consider the following scenario: Imagine one can entangle 10 atoms to reach the Heisenberg limit of 100~mrad, \textit{i.e.}, a factor of 10 reduction in noise variance with respect to the original 10 atom SQL of 320~mrad. In comparison, recent entanglement generation results in large ensembles approaching $10^6$ atoms are very far from the $10^6$ atom Heisenberg limit of 0.001~mrad, but they do realize a factor of 10 reduction in noise variance with respect to the original $10^6$ atom SQL of 1~mrad.  To achieve the same phase estimation imprecision with collections of 10~atoms, one would have to perfectly entangle each collection of 10 atoms to their Heisenberg-limited sensitivity and then successfully repeat this in parallel $10^5$ times.  It is clear that this would be daunting compared to the single operations presented here.

\section{Coherence Preserving Measurements}
Here, we will focus on making highly precise collective measurements using atoms laser-cooled and trapped inside of an optical cavity.  This approach was used  to obtain an enhancement in the phase variance  by $(\Delta\theta_{SQL}/\Delta\theta)^2= 10$  (or $10\, \log_{10}[(\Delta\theta_{SQL}/\Delta\theta)^2] =10$~dB)  relative to the standard quantum limit \cite{Squeezing_Bohnet_2014}.  More recently, we have made a preliminary observation of  $(\Delta\theta_{SQL}/\Delta\theta)^2=23(5)$. These are the directly observed enhancements with no background subtractions or corrections for imperfections or readout noise.  Thus, this is the enhancement one would expect in a Ramsey measurement relative to using a non-entangled collection of atoms, although one must be careful to consider the effects of single particle and collective dephasing and decoherence \cite{Andre2004,Sorensen13,Lukin2014}.  We report the reduction in the phase variance as this reflects the reduction in the amount of resources required to achieve the same imprecision:  23 times less Ramsey evolution time or 23 times fewer atoms.  These reduced resource requirements might be applied to increase measurement bandwidth or reduce systematic errors due to atomic collisions.

\begin{figure}[htb]
\centering
\includegraphics[width=3.5in]{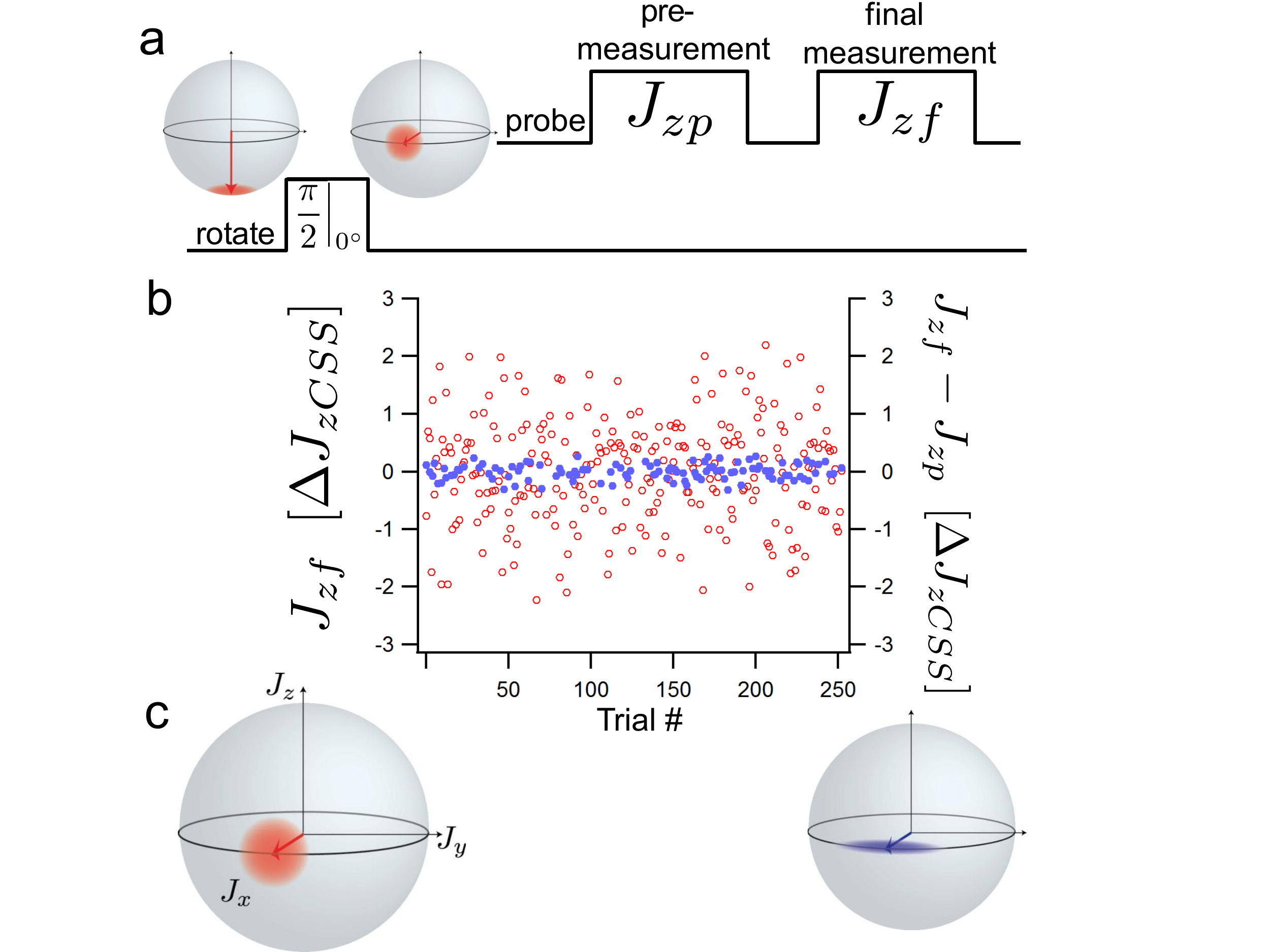}
\caption{(a) Measurement sequence for probing the atoms.  A coherent spin state is prepared by optically pumping all of the atoms into spin down. Each atom is rotated into a superposition of spin up and down, corresponding to the total spin or Bloch vector oriented along the equator.  Two consecutive measurements of the spin projection $J_z$ are then performed (100~$\mu$s each), with the pre and final measurement outcomes for a single trial labeled $J_{zp}$ and $J_{zf}$.  The sequence is then repeated many times.  (b) Measurement outcomes versus trial number. Classical rotation noise in the $\pi$/2 pulse causes classical excess noise in the observed spin noise fluctuations, so here we display simulated Gaussian noise with rms distribution equal to the predicted projection noise level fluctuations $\Delta J_{zCSS}$ (left axis, red open circles.)  The measured differential quantity $J_{zf}-J_{zp}$ (blue, right axis) shows partial cancellation of both the quantum projection noise and the excess classical noise.  The differential quantity's noise variance is 50 times below the projection noise level (or 17(1)~dB).  (c)  (left) The red data can can be visualized as arising from a fundamental blurriness of the orientation of the collective Bloch vector.  (right)  The measurement process projections the Bloch vector into a squeezed state with reduced uncertainty in the polar angle, at the expense of increased uncertainty in the azimuthal angle.  Accounting for a reduction in the size of the Bloch vector due to free space scattering and dephasing, allows a net preliminary improvement in angular variance of 23(5) or 13.7(1.0) dB below the standard quantum limit for unentangled atoms.}
\label{SubtractOut}
\end{figure}

To reduce the quantum noise, we essentially measure it and subtract it out, as shown in Fig.~\ref{SubtractOut}.  This approach was first proposed for free space probing in \cite{kuzmich98}, and our experimental and theoretical work in optical cavities is described in Refs. \cite{PhysRevLett.106.133601,Squeezing_Bohnet_2014} and \cite{PhysRevA.89.043837}.  We make a pre-measurement of the quantum spin projection noise of the total atomic ensemble, with measurement outcome labeled $J_{zp}$.   The measurement outcome is then subtracted from a final measurement of the total spin projection, with measurement outcome labeled $J_{zf}$.  The atomic projection noise cancels in the differential quantity $(J_{zf}-J_{zp})$.  If the measurement leaks no information to the environment about which atoms are in spin up versus spin down, then each atom remains in a superposition of spin up and spin down.  In a Bloch vector picture, the first measurement localizes the initially blurry quantum state into a region with less blurriness along $\hat{z}$ and thus $\theta$, at the expense of enhanced blurriness in an orthogonal spin projection.  To utilize this state to make a precise measurement, one would insert a Ramsey pulse sequence between the two measurements.

\begin{figure}[ht]
\centering
\includegraphics{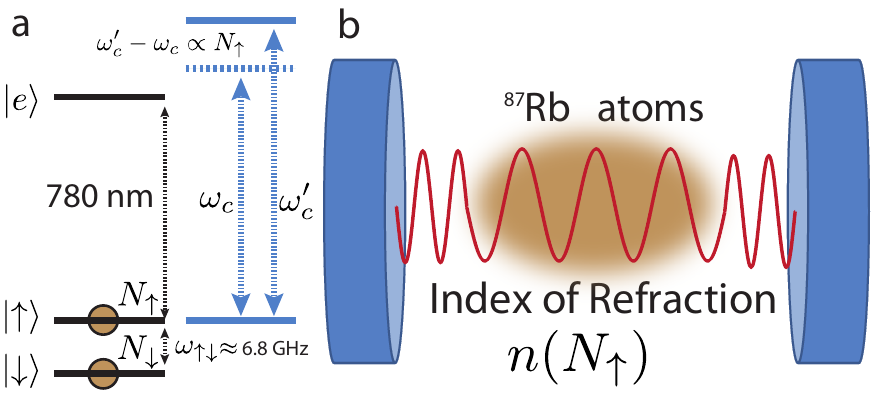}
\caption{(a) The $^{87}$Rb atomic energy levels, cavity resonance frequency $\omega_c$, and shifted resonance frequency  $\omega_c'$ when dressed by the population $N_\uparrow$ of atoms in $\ketup$. (b) The cavity frequency shift can be viewed classically as arising from a population-dependent index of refraction $n(N_\uparrow)$ that changes the wavelength of the light inside the atomic medium, causing the effective mirror separation to change.}
\label{fig:atomCavSystem}
\end{figure}

Measuring the spin projection $J_z=(N_{\uparrow} - N_{\downarrow})/2$ is achieved by attempting to count the number of atoms in spin up $N_{\uparrow}$ versus spin down $N_{\downarrow}$.  In our latest work, the effective spin-half system is composed of the two ground hyperfine states of $^{87}$Rb, $\ketup = \ket{5^2 S_{1/2}, F=2, m_F=2}$ and $\ketdown= \ket{5^2 S_{1/2}, F=1, m_F=1}$ (see Fig.~\ref{fig:atomCavSystem}). We convert the atom counting problem into a frequency measurement by placing the atoms inside of an optical cavity of finesse $F=2700$ and arranging things such that the cavity mode's resonant frequency is dispersively shifted by an amount that depends on the atomic population $N_\uparrow$. Classically, the atoms act as a medium with index of refraction $n(N_\uparrow)$.  As a result, the optical path length between the mirrors is modified, causing the cavity resonance frequency to shift.  Standard microwave $\pi$ pulses can be used to swap the populations between spin states, allowing the population $N_{\downarrow}$ to also be determined if needed.

To achieve a state-dependent shift of the cavity resonance frequency, we tune the length of the cavity such that a TEM$_{00}$ is 500 MHz blue detuned from the transition frequency between $\ketup$  and the optically excited state $\ket{e} = \ket{5^2 P_{3/2}, F'=3}$ with transition wavelength $\lambda =780$~nm.  The $\ketdown$  to $\ket{e}$ transition is much further off-resonance with the cavity mode due to the ground state hyperfine splitting of 6.8~GHz, and its interaction with the cavity can thus be largely ignored.

The cavity has a power decay rate $\kappa= 2\pi \times 3.05(5)$~MHz, a mirror separation $L=1.85(1)$~cm, a free spectral range $8.10(2)$~GHz, mirrors with radius of curvature $R_c=5$~cm, a mode waist $w_0=69~\mu$m, and mirror power transmission coefficients of $T=2010\times 10^{-6}$ and $T=130\times 10^{-6}$.  The peak single-atom Jaynes-Cummings coupling on the cycling $\ketup$ to $\ket{e}$ transition is $g=2 \pi \times 526$~kHz.  We typically load an effective atom number $N= 4 \times 10^5$ into the lattice \cite{SLV10,CBS11}.

To sense small applied rotations, it is sufficient to precisely measure changes in the population $N_\uparrow$ only (see \cite{Squeezing_Bohnet_2014} for details.)  Therefore, we concentrate on the case of making two consecutive measurements of the dressed cavity resonance frequency with measurement outcomes $f_{cp}$ and $f_{cf}$, from which we can extract a pre-measurement and final measurement of the population $N_\uparrow$.  

The goal then is to make the rms noise in the difference frequency $\Delta f_d = \Delta(f_{cf}-f_{cp})$ less than the fluctuations in the individual frequencies due to quantum projection noise $\Delta f_{PJN}$.  For the above experimental parameters we find an rms fluctuation $\Delta f_{PJN}=97$~kHz.  The experimental challenge is to achieve $\Delta f_d/\Delta f_{PJN} \ll 1$,  while also disturbing the atomic system as little as possible. In the following subsections, we provide details on our current locking scheme and characterize its performance both in terms of the technical noise contribution to $\Delta f_d$ and the associated small loss of atomic coherence.

\section{Canceling Laser-Cavity Frequency Noise}

\subsection{Experimental Scheme}
The cavity frequency shift can be easily blurred by the typical 200~kHz FWHM Lorentzian linewidth of standard external cavity diode lasers (ECDLs).  In addition, vibrations can cause the empty cavity's resonance frequency $\omega_c$ to jitter.  We use the optical lattice beam at 823~nm that is used to trap the atoms in the cavity, to also stabilize the cavity resonance frequency with a few kHz bandwidth.  In previous work, we then used DBR lasers narrowed to a few kHz to probe the laser resonance frequency \cite{Lin12}. Low frequency relative noise limited the technical noise to 17(2)~dB below the projection noise level.  Here we will describe a robust approach in which we start with  200~kHz linewidth ECDLs and demonstrate a technical noise floor on $\Delta f_d$ that is as much as 27~dB below the projection noise level.

\begin{figure}
\centering
\includegraphics[width=3.5in]{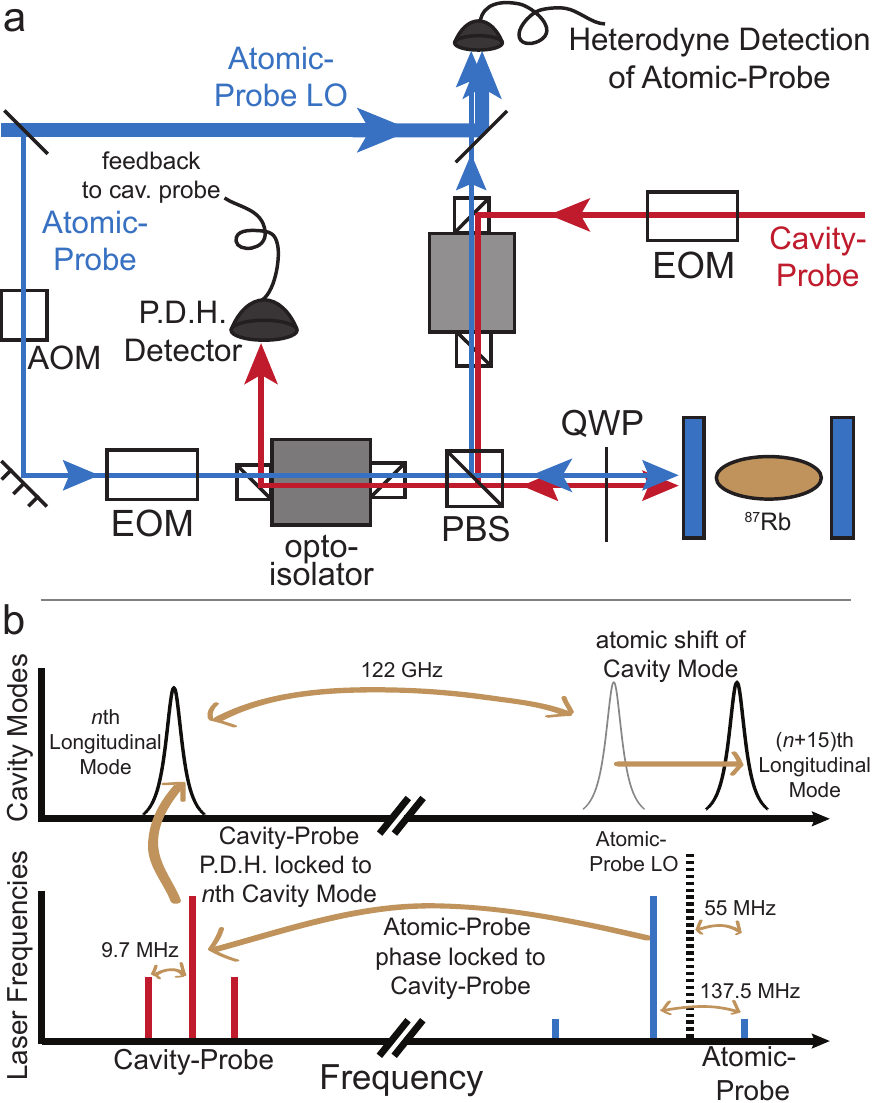}
\caption{(a) Physical scheme for probing two modes of the cavity in order to suppress probe-cavity relative frequency noise below the cavity shifts induced by quantum projection noise of the atoms. (b) The frequencies of the probed modes, the frequency components on both probes, and the several servos. The cavity-probe laser is Pound-Drever-Hall locked to one longitudinal mode of the cavity far from the atomic transition frequency. The atomic-probe laser is phase-locked to the cavity-probe and is used to probe a cavity mode close to the atomic transition frequency that is strongly tuned by an amount depending on the population $N_\uparrow$. Optical isolators, polarizing beam splitters, and a quarter wave plate are used to polarization-separate the two probes and direct the two probes to different detectors.  Electro-optic phase (EOM) and acousto-optic (AOM) modulators are used to produce sidebands for both the PDH lock and for path-length phase stabilization in the heterodyne detection of the atomic-probe.}
\label{fig:probe_cavity_scheme}
\end{figure}
Two ECDLs are used to probe two different longitudinal modes of the optical cavity.  This appraoch is related to the approach described in Ref. \cite{SLV10}. Figure~\ref{fig:probe_cavity_scheme}a depicts the physical layout used to probe the cavity. Figure~\ref{fig:probe_cavity_scheme}b outlines the various frequency components on the two probes and how the frequency chain is locked.  

The cavity-probe is tuned to a mode far from resonance with the atomic transition, while the atomic-probe is tuned to the original mode that is close to resonance with the atomic transition.  The cavity-probe laser's frequency is Pound-Drever-Hall locked using phase modulation sidebands at $f_{mv}= 9.7$~MHz. The cavity-probe light is $\sigma^-$ polarized when it hits the cavity, allowing it to be polarization-separated from the atomic-probe light that is $\sigma^+$ polarized.  The cavity-probe is detected using an avalanche photodiode with a  gain of 100. The laser is servoed to the cavity with $\sim1$~MHz unity gain frequency.

The atomic-probe laser is then phase-locked to the cavity-probe laser with unity gain frequency also close to 1~MHz.  The direct beat frequency of the two lasers is approximately 122~GHz, too high of a frequency to easily detect.  Instead, we partially bridge the frequency gap by using a fiber phase modulator driven at 13.6~GHz to place high order sidebands on light derived from the cavity-probe laser. The microwave modulation source is derived from the low phase noise 6.8~GHz local oscillator used to drive atomic rotations \cite{CBW12}.  The 9th order sideband is within 1~GHz of the atomic-probe laser and can be directly detected.  The phase of the detected signal is then phase locked to a frequency reference supplied by an AD9959 direct digital synthesizer (DDS).

%
Having established a chain to stablize the atomic-probe laser frequency relative to the optical cavity, we now consider the atomic-probe detection scheme. The atomic-probe light is reflected from the cavity and detected using a heterodyne local oscillator (LO) also derived from the atomic-probe laser.  To ensure good common mode cancellation of the atomic-probe's phase noise in the heterodyne detection, the total path lengths between the point of separation from the LO path to the point of re-overlap with the LO path are made equal to within 8~cm.  Before striking the cavity, the atomic-probe light is shifted by an acousto optic modulator (AOM) by +82.5~MHz and then phase modulated at 137.5~MHz.  This produces a small frequency sideband that is tuned to resonance with the optical cavity mode by adjusting the frequency of the DDS used to phase lock the atomic-probe laser to the cavity-probe laser. 

The frequency component that interacts resonantly with the cavity mode appears in the detected rf spectrum at 55~MHz.  The signal is IQ demodulated using phase coherent channels from the same DDS board that are used to provide the various frequency shifts and modulations.  We determine the phase of the reflected light from the two quadrature signals. The change in phase as a function of frequency detuning from the cavity resonance sets the conversion factor between detected phase and detuning.  From the measured phase noise of the reflected field, we can then estimate the relative frequency noise between the atomic-probe laser and the cavity mode that it is probing.

\begin{figure}[htb]
\centering
\includegraphics[width=.46\textwidth]{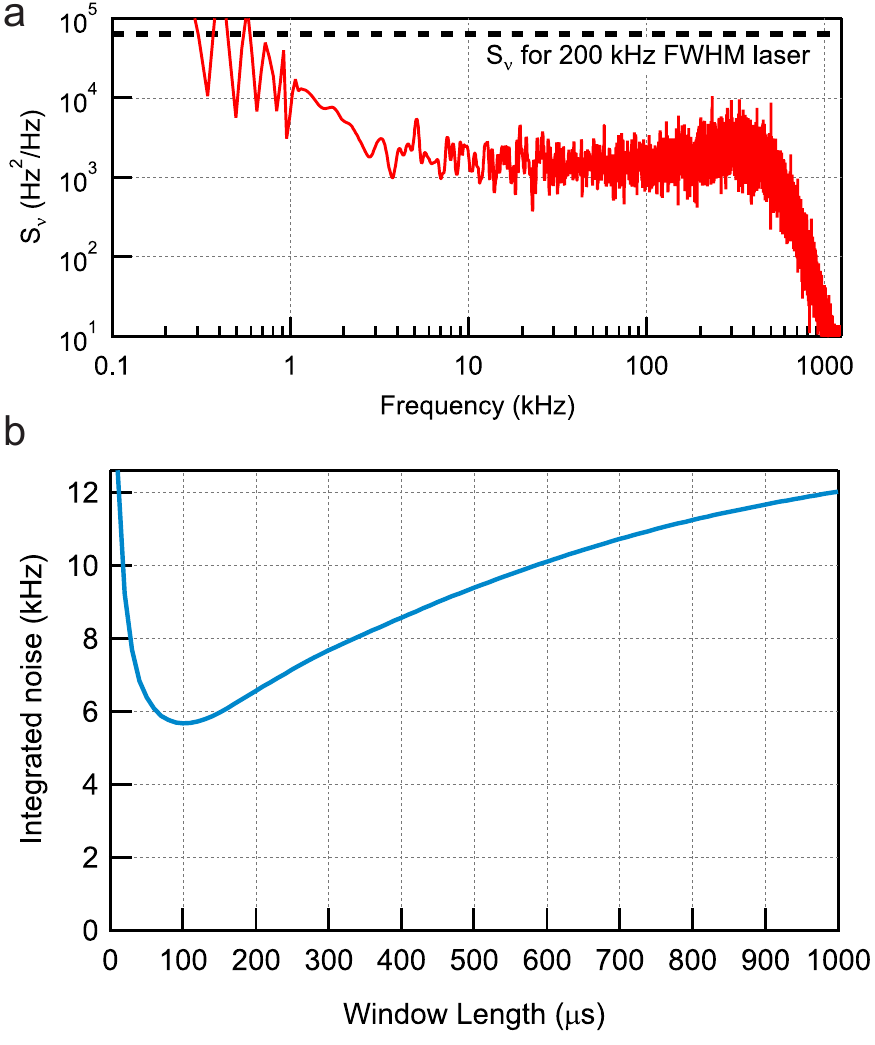}
\caption{(a) The measured power spectral density of instantaneous frequency fluctuations S$_\nu(f)$ between the atomic-probe and an empty cavity mode.  The frequency stabilization described in the text reduces the noise by close to a factor of 50 over a broad range relative to the S$_\nu$ one expects for the linewidth of our free-running 200~kHz FWHM external cavity diode lasers. For this data, the atomic and cavity-probes were set to a high enough power that increasing either did not decrease $S_\nu(f)$, so that we are sensitive only to technical noise floors.  Also, heterodyne path-length stabilization had not yet been implemented, and this is largely responsible for the rise below 2~kHz. (b) The integrated noise in the difference of two frequency measurement windows, plotted as a function of window length $T_m$, with a fixed $t=0~\mu\mathrm{s}$ window separation.}
\label{fig:varNoiseProbe2}
\end{figure}


A representative power spectral density of instantaneous frequency noise S$_\nu(f)$ is shown in Fig.~\ref{fig:varNoiseProbe2}a.  In the central flat region S$_\nu\approx1.5\times 10^3~\mathrm{Hz}^2/\mathrm{Hz}.$  This corresponds to the instantaneous frequency noise of a laser with Lorentzian FWHM $\Delta \nu= \pi\times\mathrm{S}_\nu= 5$~kHz, significantly smaller than the initial laser linewidth of 200~kHz.

The roll off at high frequency results from 300~kHz anti-aliasing low pass filters after the IQ demodulation. The rise at low frequencies is largely due to uncontrolled relative path length changes between the atomic-probe LO and atomic-probe paths.  We have found it unnecessary to stabilize this path length phase for the preliminary results presented here because it only required $200~\mu$s to measure the differential quantity $f_{cf}- f_{cp}$.

We are interested in characterizing the noise in $f_{cf}- f_{cp}$. Each measurement is the average of the measured frequency in a window of length $T_m$ and the two measurement windows have a time gap $t$ between them.  The variance is obtained by integrating $(\Delta f_d)^2=\int_0^\infty\mathrm{S}_\nu(f)\, T(f)\, \mathrm{d} f$.  The transfer function is $T(f)= 4\sin^2(\pi f (T_m+t))\sin^2(\pi f T_m)/(\pi f T_m)^2.$  Figure~\ref{fig:varNoiseProbe2}b shows the measured noise variance $\Delta f_d$ as a function of the measurement window length $T_m$ with $t=0~\mu$s.  For this data, the minimum is $\Delta f_d=6$~kHz at $T_m= 100~\mu$s.  The rise at longer times is dominated by path length fluctuations.  

The path length noise has recently been suppressed by actively stabilizing the relative path length. This is achieved using the much stronger probe component that is detuned from the cavity mode by 137.5~MHz, and produces a signal in the atomic-probe heterodyne detector at 82.5~MHz.  Appropriate demodulation allows us to derive an error signal to phase-lock the relative phase between probe path and the heterodyne reference path.  The phase is stabilized by adjusting the phase of the 82.5~MHz driving the AOM that shifts the atomic-probe. 

\subsection{Degree of coherence preservation}

The previous measurements were made at very high probe powers with the no atoms in the cavity.  As a result, the photon shot noise and  technical noise sources of the detectors were negligible compared to contributions from other technical noise sources.  However, as the power in the cavity-probe is increased, the amount of squeezing may become limited by additional scattering of light from the atoms, potentially leading to single-particle wavefunction collapse (loss of signal) and Raman transitions to other ground hyperfine states (a source of additional noise.)  Inhomogeneous differential light shifts of the spin transition frequency can also lead to dephasing that can be spin-echoed away, but perhaps imperfectly.

\begin{figure}[htb]
\centering
\includegraphics[width=0.46\textwidth]{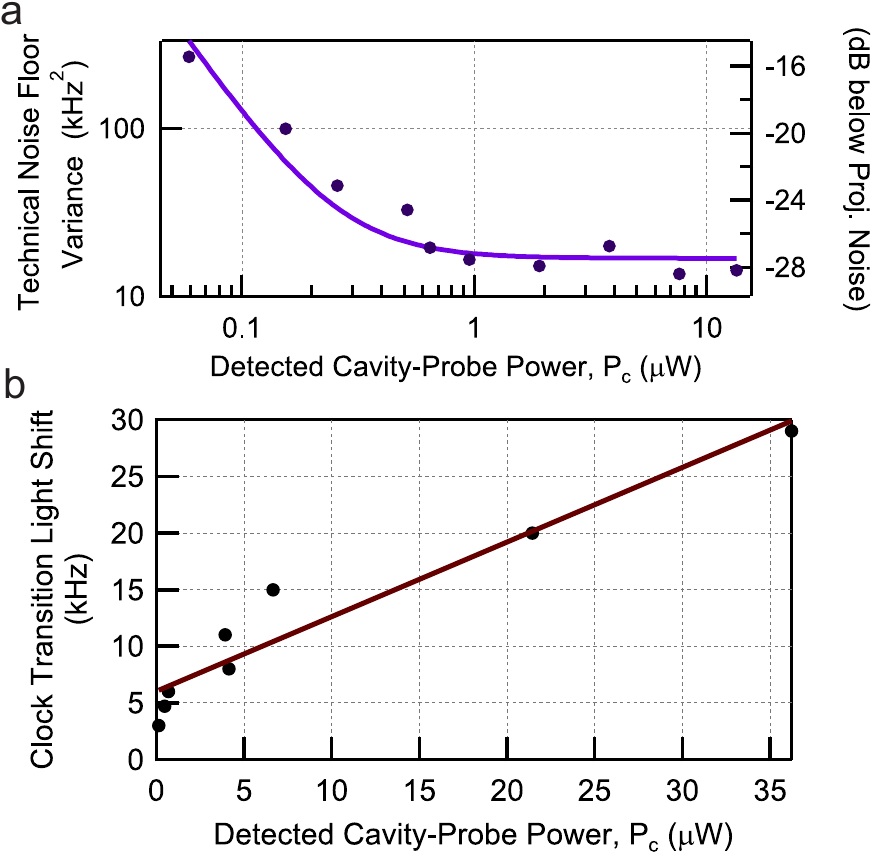}
\caption{ (a) The 2-window noise variance $(\Delta f_d)^2$ is plotted versus the detected power of the cavity-probe beam for $T_m=40~\mu$s and $t=0$.  Above 1~$\mu$W, the noise variance saturates to 27~dB below the quantum projection noise level.  (b)  The measured light shift of the $^{87}$Rb clock transition is plotted versus the detected power of the cavity-probe beam.  The shift is approximately 660~Hz per $\mu$W.  The non-zero light shift at $P_c=0$ is due to an additional constant light shift from the 823~nm optical lattice.  Due to differing Clebsch-Gordan coefficients, the shift of the $\ketdown$ to $\ketup$ transition frequency $\omega_{\uparrow\downarrow}$ is approximately 13 times larger, but still causes very little coherence loss for the preliminary squeezing results presented here.}
\label{fig:NoiseVersusCavityProbePower}
\end{figure}

Figure~\ref{fig:NoiseVersusCavityProbePower}a shows the noise variance $(\Delta f_d)^2$ between two measurements of the cavity resonance frequency versus the cavity-probe power incident on the cavity $P_c$.  Here the atomic-probe power is increased such that its photon shot noise contribution is negligible.  The right hand axis translates the noise variance into an equivalent uncertainty relative to a quantum projection noise level $\Delta f_{PJN} = 97$~kHz.  For powers above 1~$\mu$W, the technical noise floor saturates to 27~dB below the projection noise level.  For lower powers, the noise variance scales as $(\Delta f_d)^2 \propto 1/P_c^2$, indicating that lower powers $P_c$ might be utilized with improved photodetection.

The cavity-probe induces a differential light shift of the transition frequency $\omega_{\uparrow\downarrow}$ between down and up.  We measure the differential shift of the clock transition frequency $\ket{F=1, m_F=0}$ to $\ket{F=2, m_F=0}$ versus $P_c$ in Fig.~\ref{fig:NoiseVersusCavityProbePower}b.  Appropriately rescaling for transition strengths and detunings gives an average shift of $\omega_{\uparrow\downarrow}$ by 8.5~kHz per $\mu$W.  The shift is highly inhomogeneous and leads to dephasing.  However, preliminary spin-echo measurements have shown that the atomic contrast (\textit{i.e.}, length of the Bloch vector) is reduced by less than $5\%$ at $P_c=0.5~\mu$W with $40~\mu$s measurement windows.  Thus, the atomic state is not significantly decohered by the cavity stabilization presented here.

\section{Conclusion}
Coherence preserving measurements are a powerful technique for producing large amounts of entanglement in large atomic ensembles.  Here the figure of merit is the enhancement in the estimation of a quantum phase relative to the standard quantum limit.  This figure of merit is particularly compelling because it directly connects to the application of entanglement to precision measurements with atoms and ions.  The geometry used here for proof-of-principle experiments may be amenable to enhancing optical lattice clocks \cite{Bloom14,Hinkley2013} beyond the standard quantum limit and may also allow for reduced dead time due to the non-destructive nature of the readout \cite{LWL09,Westergaard2010,Kohlhaas2015}. Additionally, the scheme presented here may allow enhancements to atom interferometers\cite{AtomInfCav} used for rotation sensing \cite{GBK97}, measurements of gravitational acceleration \cite{PCC01}, or even searches for gravitational waves \cite{AtomInterfSpace}.  Because the probe light can be switched on and off, the entanglement can be generated without perturbing the atoms during the critical Ramsey phase evolution time.  Here we have presented a scheme to realize large amounts of entanglement with standard external cavity diode lasers.  Work is currently underway to improve the net quantum efficiency for detecting the atomic-probe for even greater amounts of entanglement enhancement of phase estimation.


\section*{Acknowledgment}

The authors would like to thank Matthew A. Norcia for helpful discussions.  All authors acknowledge financial support from DARPA QuASAR, ARO, NSF PFC, and NIST. K.C.C. acknowledges support from NDSEG. This work is supported by the National Science Foundation under Grant Number 1125844.  Part numbers are given as technical information only, and do not represent endorsement by NIST.

\bibliographystyle{IEEEtran}
%

\bibliography{LowNoiseProbing}

%

\end{document}